\documentclass[11pt]{article}
\usepackage{graphicx}
\title{ Abandoning Exact $SU(3)$ in Coupled-Channel Final-State Interactions
through Reggeon Exchange for $B\rightarrow \pi\pi, K\overline{K}$}

\author{{P.~\L{}ach}\thanks{plach@alf.ifj.edu.pl}\\
 {\small \it Department of Theoretical Physics, Institute of Nuclear Physics}\\
 {\small \it Radzikowskiego 152, 31-342 Krak\'ow, Poland}}

\begin{document}
\maketitle

\begin{abstract}
For weak decays $B_d^{0}\rightarrow \pi\pi$ and $K\overline{K}$
the effects of $SU(3)$ breaking in coupled-channel final-state
interaction effects  are discussed in a Regge framework. It is
shown that  $SU(3)$ breaking in the inelastic final-state
transitions dramatically affects the phases of  the isospin
$I=0,1,2$ amplitudes in the $B_d^{0}$ decays. The effect of the
singlet penguin diagram on these phases is studied. Furthermore,
on the example of the $B_d^{0}\rightarrow \pi\pi$ decays, the
dependence of $CP$ asymmetries on the size of penguin amplitude is
analyzed.
\end{abstract}


\newpage

\section{Introduction}
Final-state interaction (FSI) effects play important role in many
physical processes, and in particular in various weak decays.
These effects may significantly affect determination of
fundamental $CP$-violating parameters since extraction of the
latter requires at least some knowledge of FSI. The role of FSI in
B decays was discussed in \cite{Zen97, Don86, Wolf95}.
Unfortunately, understanding it constitutes a difficult task for
both theory and phenomenology.

Our model of coupled-channel final-state interaction is based on a
quasielastic approximation and  Regge pole methods \cite{GW99,
DGPW98, GPW98}. The basic physical idea of Regge model is that the
high energy behavior of s-channel amplitudes is determined by
"exchanges" in the crossed channel. Our model considers
rescaterings of the type: $P_iP_j \rightarrow P_kP_l$, where
$P_iP_j$ and $P_kP_l $ denote pairs of pseudoscalar mesons:
$\pi\pi$, $K\overline{K}$, $\eta\eta'$, $\eta\eta$ and
$\eta'\eta'$. The dominant exchanges in the $t$-channel are the
Pomeron ($\mathcal{P}$) and the Regge trajectories. In that
framework the coupled-channel FSI effects for $B^0_d$ weak decays
into $\pi\pi$ and $K\overline{K}$ were discussed in Refs.
\cite{Zen99}. The calculations~\cite{Zen99} were performed under
the assumption of the exchange of the $\rho$, $f_2$, $\omega$,
$a_2$ Regge trajectories, the trajectories of their $SU(3)$
partners, and the exactly $SU(3)$-symmetric Pomeron. In this paper
we analyze in some details both the influence of $SU(3)$ breaking
in the Pomeron, and the influence of singlet penguin amplitude on
the predictions of the quasi-elastic coupled-channel Regge
approach of Ref.\cite{Zen99}. If $SU(3)$ in the Pomeron is broken
and the singlet penguin is not neglected, the conclusions of
Refs.\cite{Zen99} would have to be modified.

\section{Notation}
 We use the following phase conventions for pseudoscalar mesons:
\begin{eqnarray}
&\pi^+=-u\overline{d},
\pi^0=\frac{1}{\sqrt{2}}(u\overline{u}-d\overline{d}),
\pi^-=d\overline{u},&\nonumber\\
&\eta=\frac{1}{\sqrt{3}}(u\bar{u}+d\bar{d}-s\bar{s}),
\eta'=\frac{1}{\sqrt{6}}(u\bar{u}+d\bar{d}+2s\bar{s}),&\nonumber\\
&K^+=u\overline{s}, K^0=d\overline{s}, K^-=s\overline{u},
\overline{K}^0=-s\overline{d}.&
\end{eqnarray}

For Cabibbo-suppressed $B_d^0$ decays there are nine possible
final states composed of two pseudoscalar mesons. In the basis of
definite isospin $I$ the symmetrized two-boson states
$|(P_kP_l)_I\rangle$ are:
\begin{eqnarray}
|(\pi\pi)_{2}\rangle&=&\frac{1}{\sqrt{6}}(\pi^{+}\pi^{-}+\pi^{-}\pi^{+}+2\pi^{0}\pi^{0}),\nonumber\\
|(K\overline{K})_{1}\rangle&=&\frac{1}{2}(K^+K^- +K^-K^++K^0\overline{K^0}+\overline{K^0}K^{0}),\nonumber\\
|(\pi^0\eta)_1\rangle&=&\frac{1}{\sqrt{2}}(\pi^0\eta+\eta\pi^0),\nonumber\\
|(\pi^0\eta')_1\rangle&=&\frac{1}{\sqrt{2}}(\pi^0\eta'+\eta'\pi^0),\nonumber\\
|(\pi\pi)_{0}\rangle&=&\frac{1}{\sqrt{3}}(\pi^{+}\pi^{-}+\pi^{-}\pi^{+}-\pi^{0}\pi^{0}),\nonumber\\
|(K\overline{K})_{0}\rangle&=&\frac{1}{2}(K^+K^-+K^-K^+-K^0\overline{K^0}-\overline{K^0}K^{0}),\nonumber\\
|(\eta\eta)_0\rangle&=&\eta\eta,\nonumber\\
|(\eta\eta')_0\rangle&=&\frac{1}{\sqrt{2}}(\eta\eta'+\eta'\eta),\nonumber\\
|(\eta'\eta')_0\rangle&=&\eta'\eta'. \label{states}
\end{eqnarray}

\section{Quark Diagram Amplitudes}
\begin{figure}[h]
\begin{center}
\includegraphics[width=0.4\textwidth]{{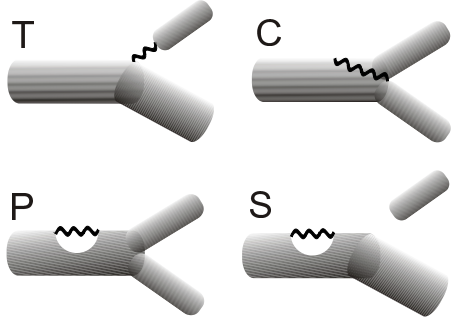}}
\end{center}
\caption{\small Graphs describing  invariant $SU(3)$-flavor
amplitudes for the decays of B mesons to a pair of light
pseudoscalar mesons. ($T$)~"Tree"; ($C$)~"Color-suppressed";
($P$)~"Penguin"; ($S$)~Additional penguin involving
flavor-$SU(3)$-singlet.
\label{4d}}
\end{figure}

The decays of $B_d^0$ mesons to two pseudoscalar mesons ($P_iP_j$)
are described by 7 flavor-$SU(3)$ invariant
amplitudes~\cite{Zen02}, but only 4 of them (Fig.~\ref{4d}):
"tree" ($T$), "color-suppressed" ($C$), "penguin" ($P$) and
additional penguin involving flavor-$SU(3)$-singlet ($S$)
diagrams, are important \cite{GR99}. We assume that
$|C|=|T|/3|r|$, with $r\approx-3$ \cite{Zen99},
$|P|\approx(0.2\div0.5) |T|$ \cite{Tev} and $|S|\approx(0.6\pm0.2)
|P|$ \cite{Tev}.

Short-distance amplitudes: $T$, $C$, $P$ and $S$, have weak and
strong phases. We can write:
\begin{eqnarray}
T&=&|T| e^{i(\gamma+\delta_T)},\nonumber\\
C&=&|C|e^{i(\pi+\gamma+\delta_C)},\mbox{for } r<0,\nonumber\\
P&=&|P| e^{i(-\beta+\delta_P)},\nonumber\\
S&=&|S|e^{i(-\beta+\delta_S)}, \label{4diagrams}
\end{eqnarray}
where $\beta$, $\gamma$, ($\delta$) are weak (strong) phases. It
is possible that the short-distance weak amplitudes have large
strong phases~\cite{L&Rosner}. However, since we want to study FSI
we neglect these phases (i.e. we set $\delta_T, \delta_C,
\delta_P, \delta_S=0$). For the weak phases we assume \cite{Tev}
\begin{eqnarray}
\gamma&=&(60.0^{+5.4}_{-6.8})^o,\nonumber\\
\beta&=&(22.2\pm2.0)^o. \label{gamma&beta}
\end{eqnarray}
 Furthermore, we neglect the
electroweak penguins diagrams~\cite{Rosner}. In terms of quark
diagram amplitudes the weak decays of $B^0_d$ to the states
defined in Eq.~\ref{states} are given by:
\begin{eqnarray}
\langle(\pi\pi)_2|w|B_d^0\rangle&=&-\frac{1}{\sqrt{6}}(T+C),\nonumber\\
\langle(K\overline{K})_1|w|B_d^0\rangle&=&-\frac{1}{2}P,\nonumber\\
\langle(\pi^0\eta)_1|w|B_d^0\rangle&=&-\frac{1}{2\sqrt{3}}(2P+S),\nonumber\\
\langle(\pi^0\eta')_1|w|B_d^0\rangle&=&-\frac{1}{\sqrt{6}}(P+2S),\nonumber\\
\langle(\pi\pi)_0|w|B_d^0\rangle&=&-\frac{1}{2\sqrt{3}}(2T-C+3P),\nonumber\\
\langle(K\overline{K})_0|w|B_d^0\rangle&=&\frac{1}{2}P,\nonumber\\
\langle(\eta\eta)_0|w|B_d^0\rangle&=&\frac{1}{3}(C+P+S),\nonumber\\
\langle(\eta\eta')_0|w|B_d^0\rangle&=&\frac{1}{6}(2C+2P+5S),\nonumber\\
\langle(\eta'\eta')_0|w|B_d^0\rangle&=&\frac{1}{6}(C+P+4S).
\label{diagrams}
\end{eqnarray}
We assumed $SU(3)$ symmetry in weak decays, i.e. equal amplitudes
for the production of strange ($s\overline{s}$) and nonstrange
quark pairs.

\section{Final State Interaction}
\subsection{General Framework}
The weak amplitude $w$ is changed by isospin-conserving strong
interaction $S_{FSI}$ in the final state~\cite{Zen99} into a
FSI-corrected weak amplitude $W$:
\begin{equation}
(B_d^0\stackrel{w}{\rightarrow}
(P_iP_j)_I\stackrel{S_{FSI}}{\longrightarrow}(P_kP_l)_I)\equiv
B_d^0\stackrel{W}{\Longrightarrow} (P_kP_l)_I,
\end{equation}
where subscript $I$ denotes isospin. We describe $S_{FSI}$ in the
Regge pole model as used in \cite{GW99}-\cite{Zen99}. In the
energy range $s\simeq{m_{B_d}}^2=27.88$~GeV$^2$ the Pomeron
($\mathcal{P}$) contribution to the $t$-channel amplitude is
phenomenologically well described by the formula \cite{GPW98}
\begin{equation}
A_{\mathcal{P}}(P_iP_j)=i\beta_{\mathcal{P}}^{P_i}\beta_{\mathcal{P}}^{P_j}e^{i(b_{\mathcal{P}}^{P_i}+b_{\mathcal{P}}^{P_j})t}s,
\end{equation}
where the residue
$\beta_{\mathcal{P}}^{P_i}\beta_{\mathcal{P}}^{P_j}$ and slope
$b_{\mathcal{P}}^{P_i}+b_{\mathcal{P}}^{P_j}$ depend on the
scattering process considered. Calculations of the s-channel $l=0$
waves $a_\mathcal{P}(P_iP_j)$, give, for the Pomeron \cite{GPW98}:
\begin{equation}
a_{\mathcal{P}}(P_iP_j)=i\mathcal{P}_{P_iP_j}=i\frac{\beta_{\mathcal{P}}^{P_i}\beta_{\mathcal{P}}^{P_j}}{b_{\mathcal{P}}^{P_i}+b_{\mathcal{P}}^{P_i}}.
\label{a_p}
\end{equation}
From \cite{Ser64, Phi71} we obtain
\begin{eqnarray}
\beta_{\mathcal{P}}^{\pi}&=&3.48 \mathrm{\sqrt{mb}},\\
\beta_{\mathcal{P}}^{K}&=&2.74 \mathrm{\sqrt{mb}}
\end{eqnarray}
and
\begin{eqnarray}
b_{\mathcal{P}}^{\pi}&=&2.06 \mathrm{GeV^{-2}},\\
b_{\mathcal{P}}^{K}&=&0.8 \mathrm{GeV^{-2}}.
\end{eqnarray}
The simple relations between $\beta_{\mathcal{P}}^{\eta(\eta')}$,
$b_{\mathcal{P}}^{\eta(\eta')}$ and
$\beta_{\mathcal{P}}^{\pi(K)}$, $b_{\mathcal{P}}^{\pi(K)}$ for
 broken $SU(3)$ are given by
\begin{eqnarray}
\beta_{\mathcal{P}}^{\eta}&=&\frac{1}{3}(\beta_{\mathcal{P}}^{\pi}+2\beta_{\mathcal{P}}^{K}),\\
\beta_{\mathcal{P}}^{\eta'}&=&\frac{1}{3}(-\beta_{\mathcal{P}}^{\pi}+4\beta_{\mathcal{P}}^{K}),
\end{eqnarray}
and
\begin{eqnarray}
b_{\mathcal{P}}^{\eta}&=&\frac{1}{3}(b_{\mathcal{P}}^{\pi}+2b_{\mathcal{P}}^{K}),\\
b_{\mathcal{P}}^{\eta'}&=&\frac{1}{3}(-b_{\mathcal{P}}^{\pi}+4b_{\mathcal{P}}^{K}).
\label{b_p}
\end{eqnarray}
From Eqs.~\ref{a_p}~--~\ref{b_p} we find:
\begin{eqnarray}
\label{pomeron}
\mathcal{P}_{\pi\pi}=& 2.9&  \mathrm{mb \,GeV^2},\nonumber\\
\mathcal{P}_{K\overline{K}}=& 4.9&  \mathrm{mb \,GeV^2},\nonumber\\
\mathcal{P}_{\pi\eta}= &3.2&  \mathrm{mb \,GeV^2},\nonumber\\
\mathcal{P}_{\pi\eta'}=& 3.6&  \mathrm{mb \,GeV^2},\nonumber\\
\mathcal{P}_{\eta\eta}=& 3.7&  \mathrm{mb \,GeV^2},\nonumber\\
\mathcal{P}_{\eta'\eta}=& 4.8&  \mathrm{mb \,GeV^2},\nonumber\\
\mathcal{P}_{\eta'\eta'}= &8.7&  \mathrm{mb \,GeV^2}.
\end{eqnarray}
In the $SU(3)$ symmetric case we have
$\mathcal{P}_{P_iP_j}=\mathcal{P}=3.6$~mb\,GeV$^2$.

Many authors restrict their studies to elastic rescattering only.
In Regge language this is described in terms of a Pomeron
exchange. But at $s=m_B^2$ contributions from other inelastic
nonleading Regge exchanges are not completely negligible
\cite{Zen99}.
\begin{figure}[h]
\begin{center}
\includegraphics[width=0.38\textwidth]{{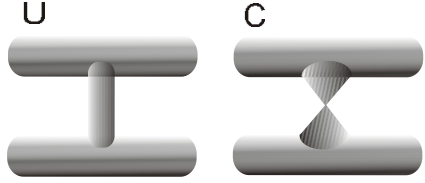}}
\end{center}
\caption{\small FSI diagrams ($\textbf{U}$) Uncrossed Reggeon
exchange, ($\textbf{C}$) Crossed Reggeon exchange.
\label{CU}}
\end{figure}
There are two types of  contributions from exchange-degenerate
Reggeons corresponding to two different diagrams (crossed
$\textbf{C}$ and uncrossed $\textbf{U}$, see Fig.~\ref{CU}). The
contributions of diagrams $\textbf{U}$ and $\textbf{C}$ differ in
their phases \cite{Zen99}.The calculations of the $s$-channel
$l=0$ partial waves amplitudes $a_U$ for uncrossed Reggeon
exchange and $a_C$ for crossed Reggeon exchange give \cite{GPW98}:
\begin{equation}
a_U=-\frac{R}{\alpha'}\frac{i(\frac{s}{s_0})^{-1/2}(\ln {
\frac{s}{s_0}}+i\pi)}{\ln^2{\frac{s}{s_0}}+\pi^2}
\end{equation}
and
\begin{equation}
a_C=\frac{R}{\alpha'}\frac{(\frac{s}{s_0})^{-1/2}}{\ln{\frac{s}{s_0}}},
\end{equation}
where $R$ is the Regge residue fitted from experiment
\cite{Zen99},\cite{Phi71} :
\begin{equation}
R=-4g^2(\omega,KK)=-\frac{4}{9}g^2(\omega,pp)=-13.1\,\,\mathrm{mb},
\end{equation}
and
\begin{equation}
\alpha'\approx1 \,\,\mathrm{GeV^{-2}}.
\end{equation}
The scale factor $s_0$ is taken as $1 \mathrm{GeV^2}$.

Inelastic FSI means here the coupled-channel effects of the type:
$\pi\pi\rightarrow K\overline{K}$, $\eta_8\eta_8$, $\eta_1\eta_8$,
\dots, and $K\overline{K} \rightarrow \pi\pi$, $\eta_8\eta_8$,
$\eta_1\eta_8$, \dots etc in the final state. Inclusion of such
processes was shown in \cite{Zen99} to be very important. There
are tree separate non-communicating FSI sectors of different
isospin $(I=0,1,2)$.

In the $I=2$ sector one obtains only the contribution from the
crossed diagram of Fig.~\ref{CU}:
\begin{equation}
\textbf{U}_2=[\langle
(\pi\pi)_2|\textbf{U}_2|(\pi\pi)_2\rangle]=0,
\end{equation}
\begin{equation}
\textbf{C}_2=[\langle
(\pi\pi)_2|\textbf{C}_2|(\pi\pi)_2\rangle]=2.
\end{equation}

In the $I=1$ sector there are three states, and consequently we
have coupled-channel effects described together with quasi-elastic
effects by two $2\times2$ matrices. One obtains:
\begin{equation}
\textbf{U}_1=[\langle i|\textbf{U}_1|j\rangle]
 = \left[
\begin{array}{ccc}
\epsilon^2 &  \frac{2}{\sqrt{3}}\epsilon & \sqrt{\frac{2}{3}}\epsilon \\
\frac{2}{\sqrt{3}}\epsilon & \frac{4}{3} & \frac{2}{3}\sqrt{2} \\
\sqrt{\frac{2}{3}}\epsilon & \frac{2}{3}\sqrt{2} &  \frac{2}{3}
\end{array}
\right]
\end{equation}
and
\begin{equation}
 \textbf{C}_1=[\langle i|\textbf{C}_1|j\rangle]\\
 = \left[
\begin{array}{ccc}
0 &  -\frac{2}{\sqrt{3}}{|\epsilon|} & 2\sqrt{\frac{2}{3}}|\epsilon| \\
-\frac{2}{\sqrt{3}}{|\epsilon|} & \frac{4}{3} & \frac{2}{3}\sqrt{2} \\
2\sqrt{\frac{2}{3}}|\epsilon| & \frac{2}{3}\sqrt{2} &  \frac{2}{3}
\end{array}
\right].
\end{equation}
The states in the rows and columns are (from top to bottom and
from left to right): $i,j= |(K\overline{K})_1\rangle$,
$|(\pi^0\eta)_1\rangle$ and $|(\pi^0\eta')_1\rangle$.

In the $I=0$ sector there are five states with rows and columns
corresponding to the states (from top to bottom and from left to
right): $i,j= |(\pi\pi)_0\rangle$, $|(K\overline{K})_0\rangle$,
$|(\eta\eta)_0\rangle$, $|(\eta\eta')_0\rangle$ and
$|(\eta'\eta')_0\rangle$. One obtains:\\
 $\textbf{U}_0=[\langle i|\textbf{U}_0|j\rangle=$
\begin{equation}
 =\left[
\begin{array}{ccccc}
3 &  -\sqrt{3}\epsilon & -\frac{2}{\sqrt{3}} & -\frac{2}{\sqrt{3}} &-\frac{1}{\sqrt{3}} \\
-\sqrt{3}\epsilon &  \epsilon^2+2 & \frac{4}{3}\epsilon &-\frac{2}{3}\epsilon & \frac{5}{3}\epsilon\\
-\frac{2}{\sqrt{3}} &  \frac{4}{3}\epsilon & \frac{2}{9}(2+\epsilon^2) & \frac{4}{9}(1-\epsilon^2) & \frac{2}{9}(1+2\epsilon^2)\\
-\frac{2}{\sqrt{3}} &  -\frac{2}{3}\epsilon & \frac{4}{9}(1-\epsilon^2) & \frac{4}{9}(1+2\epsilon^2) & \frac{2}{9}(1-4\epsilon^2)\\
-\frac{1}{\sqrt{3}} & \frac{5}{3}\epsilon &
\frac{2}{9}(1+2\epsilon^2) & \frac{2}{9}(1-4\epsilon^2) &
\frac{1}{9}(1+8\epsilon^2)
\end{array}
\right]
\end{equation}
and  $\textbf{C}_0=[\langle i|\textbf{C}_0|j\rangle]=$
\begin{equation}
 = \left[
\begin{array}{ccccc}
-1 &  0 & -\frac{2}{\sqrt{3}} & -\frac{2}{\sqrt{3}} &-\frac{1}{\sqrt{3}} \\
0 & 0 & -\frac{4}{3}|\epsilon| &\frac{2}{3}|\epsilon| & \frac{4}{3}|\epsilon|\\
-\frac{2}{\sqrt{3}} &  -\frac{4}{3}|\epsilon| & \frac{2}{9}(2+1|\epsilon|^2) & \frac{4}{9}(1-|\epsilon|^2) & \frac{2}{9}(1+2|\epsilon|^2)\\
-\frac{2}{\sqrt{3}} &  \frac{2}{3}|\epsilon| & \frac{4}{9}(1-|\epsilon|^2) & \frac{4}{9}(1+2|\epsilon|^2) & \frac{2}{9}(1-4|\epsilon|^2)\\
-\frac{1}{\sqrt{3}} &  \frac{4}{3}|\epsilon| &
\frac{2}{9}(1+2|\epsilon|^2) & \frac{2}{9}(1-4|\epsilon|^2) &
\frac{1}{9}(1+8|\epsilon|^2)
\end{array}
\right].
\end{equation}
The parameter $\epsilon$ ($\epsilon^2$) describes suppression of
propagation of one (two) strange quarks in the $t$-channel. For
the $SU(3)$ discussion of coupled-channel effects
$\epsilon=1$ \cite{Zen99}. A more realistic assumption used in this paper is:\\
\begin{equation}
\epsilon=(-\frac{s}{s_0})^{\alpha_0(K^*) - \alpha_0(\rho)}\approx
0.5e^{-i 36^o},
\label{epsilon}
\end{equation}
where $\alpha_0(K^*)\approx 0.3$ and $\alpha_0(\rho)\approx 0.5$
are Reggeon's parameters.

Let us now connect weak decays and strong interactions in the
final state. We can obtain amplitudes
$\langle(P_iP_j),I|W|B_d^0\rangle $ of $B_d^0$ decay  to states
$(P_iP_j)_I$ from:
\begin{eqnarray}
\label{main1}
&&\langle(P_iP_j),I|W|B_d^0\rangle=\langle(P_iP_j),I|S^{1/2}_{FSI}w|B_d^0\rangle=\nonumber\\&&=\sum_{\textbf{V}}\langle(P_iP_j),I|\textbf{V},I\rangle\langle\textbf{V},I|S^{1/2}_{FSI}|\textbf{V},I\rangle\langle
(P_iP_j),I |w|B_d^0\rangle,
\end{eqnarray}
with $\langle(P_iP_j),I|\textbf{V},I\rangle$ are eigenvectors for
$S_{FSI}^{1/2}(I)=iP+a_UU_I+a_CC_I$ matrices. We assume now that
the FSI-corrected weak decay amplitudes differ from quark-level
expressions Eq.~\ref{diagrams} by hadronic phase factors only
($S_{FSI}=e^{2i\delta}$) \cite{GPW98},\cite{Zen99}.

\subsection{Numerical Results }
Using Eq.~\ref{main1} one obtains the numbers given in the
right-hand side of Tables~\ref{TAB1}, \ref{TAB2}  and in
Table~\ref{TAB3}. For the sake of comparison,in the left-hand side
of Table~\ref{TAB1}, \ref{TAB2} we added amplitude phases with
$SU(3)$ symmetric FSI (Table~\ref{TAB1}), as well as amplitude
phases calculated without FSI effects (Table~\ref{TAB2}).

\begin{table}[h]
\caption{\small Comparison of calculated values of amplitude
phases for $B_d^0$ decays with weak phases set to $0$.}
\label{TAB1}
\begin{center}
{\footnotesize
\begin{tabular}{|l||c|c c|c c|c c|c c|}
 \hline
&\multicolumn{1}{c|}{No c. c. }&\multicolumn{8}{c|}{Coupled channels (c. c.)}\\
\cline{2-10}
&\multicolumn{5}{c|}{$\epsilon=1$}&\multicolumn{4}{c|}{$\epsilon=0.5e^{-i36^o}$ }\\
\cline{2-10}
\multicolumn{1}{|c||}{Phase $\varphi$,}&\multicolumn{3}{c|}{Ref.\cite{Zen99},}&\multicolumn{4}{c|}{$SU(3)$broken}&\multicolumn{2}{c|}{$\mathcal{P}=$}\\
\multicolumn{1}{|c||}{$\varphi\in(-180^o,180^o)$}&\multicolumn{3}{c|}{$\mathcal{P}=3.6$mbGeV$^2$}&\multicolumn{4}{c|}{in Pomerons}&\multicolumn{2}{c|}{$3.6$ mbGeV$^2$}\\
\cline{2-10}
&\multicolumn{1}{c|}{}&\multicolumn{8}{c|}{$|P|/|T|=$}\\
& &\multicolumn{1}{c}{$0.04$} & \multicolumn{1}{c|}{$0.2$} & \multicolumn{1}{c}{$0.04$}&\multicolumn{1}{c|}{$0.2$}& \multicolumn{1}{c}{$0.04$}&\multicolumn{1}{c|}{$0.2$}& \multicolumn{1}{c}{$0.04$}&\multicolumn{1}{c|}{$0.2$}\\
\hline \hline
$\varphi_{\pi\pi}^2$                                       &$112^o$ & $112^o$ & $112^o$ & $117^o$ & $117^o$    &$117^o$ & $117^o$  &$112^o$ & $112^o$       \\
$\varphi_{\pi\pi}^0$                                       & $94^o$ & $93^o$  & $94^o$  &$85^o$  & $89^o$     & $89^o$  & $91^o$ & $92^o$  & $94^o$        \\
$\varphi_{\pi\pi }^2-\varphi _{\pi\pi }^0$                 &$18^o$ & $19^o$  & $18^o$  &$32^o$  & $28^o$       & $28^o$  & $26^o$ & $20^o$  & $18^o$      \\
$\varphi_{K\bar{K}}^1$                                     & $95^o$  & $85^o$  & $85^o$  &$104^o$  & $103^o$   & $91^o$  & $91^o$  & $93^o$  & $93^o$      \\
$\varphi_{K\bar{K}}^0$                                    & $103^o$ & $168^o$ & $137^o$ &$168^o$ & $123^o$    & $110^o$ & $98^o$ & $123^o$ & $113^o$    \\
$\varphi_{K\overline{K}}^1-\varphi _{K\overline{K}}^0$      &$-8^o$ & $-83^o$ & $-52^o$ & $-59^o$ & $-20^o$      &$-19^o$ & $-7^o$  &$-30^o$ & $-20^o$    \\
 \hline
\end{tabular}
}
\end{center}
\end{table}
In order to make comparison with \cite{Zen99} possible, we first
put the phases $\beta$ and $\gamma$ to zero. For this case, in
Table~\ref{TAB1} we present the dependence of amplitude phases on
$SU(3)$ breaking in the Pomeron coupling ($\mathcal{P}_{P_iPj}$)
and through the parameter $\epsilon$, and on the combination of
these two effects. It is interesting to see where $SU(3)$ breaking
is important for numerical results. If we switch $SU(3)$ breaking
on in Pomerons only, and compare with \cite{Zen99} (left-hand side
of Table~\ref{TAB1}) we obtain for $|P|/|T|=0.2$:
$\varphi_{\pi\pi}^2-\varphi_{\pi\pi}^0=18^o\rightarrow28^o$ and
$\varphi_{K\overline{K}}^1-\varphi_{K\overline{K}}^0=-52^o\rightarrow-20^o$.
The effect is large. If we switch  $SU(3)$ breaking on only  in
$\epsilon$ ($\mathcal{P}_{P_iPj}=3.6$ mbGeV$^2$), we obtain:
$\varphi_{\pi\pi}^2-\varphi_{\pi\pi}^0=18^o\rightarrow18^o$ and
$\varphi_{K\overline{K}}^1-\varphi_{K\overline{K}}^0=-52^o\rightarrow-20^o$.
We see that in this case we may neglect the effect of $SU(3)$
breaking in $(\pi\pi)_I$ phases, but in $(K\overline{K})_I$ phases
the effect is large. Now, we combine both effects. Comparing
appropriate columns we see that:
$\varphi_{\pi\pi}^2-\varphi_{\pi\pi}^0=18^o\rightarrow26^o$ and
$\varphi_{K\overline{K}}^1-\varphi_{K\overline{K}}^0=-52^o\rightarrow-7^o$
for $|P|/|T|=0.2$\,. We see that in $(K\overline{K})_I$ both
effects are important, and neither of them can be neglected.
\begin{table}[h]
\caption{\small Comparison of calculated values of phase shifts
for $B_d^0$ decays in case of nonzero weak phases.} \label{TAB2}
\begin{center}
{\footnotesize
\begin{tabular}{|l||c c c|c c c|c c c|}
 \hline
&\multicolumn{3}{c|}{No FSI}&\multicolumn{6}{c|}{Coupled channels
with $SU(3)$ breaking}
\\\cline{2-10}
\multicolumn{1}{|c||}{Phase $\varphi$,}&\multicolumn{6}{c|}{$S=0$}&\multicolumn{3}{c|}{$|S|/|P|=0.6$}\\
\cline{2-10}
\multicolumn{1}{|c||}{$\varphi\in(-180^o,180^o)$}&\multicolumn{9}{c|}{$|P|/|T|=$}\\
&\multicolumn{1}{c}{$0.04$} &\multicolumn{1}{c}{ $0.2$} &\multicolumn{1}{c|}{ $0.35$}&\multicolumn{1}{c}{$0.04$}&\multicolumn{1}{c}{$0.2$}& \multicolumn{1}{c|}{$0.35$}& \multicolumn{1}{c}{$0.2$} & \multicolumn{1}{c}{$0.35$} &\multicolumn{1}{c|}{ $0.5$}\\
 \hline \hline
$\varphi_{\pi\pi}^2$                                 &  $-120^o$  & $-120^o$  & $-120^o$   & $-3^o$   & $-3^o$  & $-3^o$        & $-3^o$        & $-3^o$     & $-3^o$ \\
$\varphi_{\pi\pi}^0$                                 &  $-123^o$  & $-135^o$  & $-145^o$   & $-39^o$  & $-49^o$ & $-56^o$       & $-47^o$       & $-53^o$    & $-58^o$\\
$\varphi_{\pi\pi }^2-\varphi _{\pi\pi }^0$           & $3^o$      & $15^o$    & $25^o$     & $36^o$   & $46^o$  & $53^o$        & $44^o$        & $50^o$     & $55^o$\\
$\varphi_{K\bar{K}}^1$                               &  $158^o$   & $158^o$   & $158^o$    & $-98^o$ & $-98^o$    & $-98^o$      & $-84^o$      & $-84^o$   & $-84^o$\\
$\varphi_{K\bar{K}}^0$                               & $-22^o$    & $-22^o$   & $-22^o$    &$104^o$   & $81^o$  & $80^o$        & $94^o$        & $91^o$     & $90^o$\\
$\varphi_{K\overline{K}}^1-\varphi_{K\overline{K}}^0$& $180^o $   & $180^o$   & $180^o$    & $158^o$  & $179^o$ & $178^o$       & $-178^o$       & $-175^o$    & $-174^o$\\
 \hline
\end{tabular}
}
\end{center}
\end{table}

The numbers given in  Table~\ref{TAB2} are obtained with realistic
weak phases of Eq.~\ref{gamma&beta}. Comparing appropriate columns
in Table~\ref{TAB2} we see that inclusion of weak phases and
coupled-channel effects change amplitude phases in the considered
model: $\varphi^2_{\pi\pi}-\varphi^0_{\pi\pi}=25^o\rightarrow
53^o$ and
$\varphi^1_{K\overline{K}}-\varphi^0_{K\overline{K}}=180^o\rightarrow178^o$
for $|P|/|T|=0.35$. Amplitude phases strongly depend on the ratio
$|P|/|T|$, for instance: $\varphi_{K\overline{K}}^0(|P|/|T|=0.35,
S=0)-\varphi_{K\overline{K}}^0(|P|/|T|=0.04, S=0)=-24^o$. However
in the region $|P|/|T|\in(0.2, 0.5)$ the dependence is not strong
(no more than $7^o$).

\begin{table}[h]
\caption{\small Influence of the singlet penguin on phase shifts
in $B_d^0$ decays} \label{TAB3}
\begin{center}
{\footnotesize
\begin{tabular}{|l||c c c|c c c|}
 \hline
\multicolumn{1}{|c||}{}&\multicolumn{3}{c|}{$S=0$}&\multicolumn{3}{c|}{$|S|=0.6|P|$}\\
\cline{2-7}
\multicolumn{1}{|c||}{Phase $\varphi$,}&\multicolumn{6}{c|}{$|P|/|T|=$}\\
$\varphi\in(-180^o,180^o)$&\multicolumn{1}{c}{$0.2$}& \multicolumn{1}{c}{$0.35$}& \multicolumn{1}{c|}{$0.5$} & \multicolumn{1}{c}{$0.2$} & \multicolumn{1}{c}{$0.35$} & \multicolumn{1}{c|}{$0.5$}\\
 \hline \hline
$\varphi_{\pi^0\eta}^1$              & $-95^o$   & $-95^o$   &$-95^o$ & $-82^o$    & $-90^o$  & $-90^o$\\
$\varphi_{\pi^0\eta'}^1$             & $-70^o$   & $-70^o$   &$-70^o$ & $-86^o$    & $-86^o$  & $-86^o$\\
$\varphi_{\eta\eta}^0$               & $-127^o$  & $-160^o$  &$157^o$  &$-140^o$   & $136^o$   & $109^o$\\
$\varphi_{\eta\eta'}^0$              & $-133^o$  & $-179^o$  &$148^o$  & $140^o$   & $105^o$   & $100^o$\\
$\varphi_{\eta'\eta'}^0$             & $160^o$    &$115^o$    &$114^o$   & $74^o$    & $79^o$  & $81^o$\\
\hline
\end{tabular}
}
\end{center}
\end{table}

Now we discuss the influence of the singlet penguin.  From
Table~\ref{TAB2} we see that the phases for decays
$B_d^0\rightarrow \pi\pi , K\overline{K}$ do not depend very
strongly on the inclusion of the singlet penguin, for instance:
$\varphi_{\pi\pi}^0(|P|/|T|=0.35,
|S|/|P|=0.6)-\varphi_{\pi\pi}^0(|P|/|T|=0.35, S=0)=-3^o$,
$\varphi_{K\overline{K}}^0(|P|/|T|=0.35,
|S|/|P|=0.6)-\varphi_{K\overline{K}}^0(|P|/|T|=0.35, S=0)=11^o$.
Thus, for $B^0_d\rightarrow\pi\pi,K\overline{K}$ the effect of the
singlet penguin may be neglected. However, from Table~\ref{TAB3}
we see that the influence of the singlet penguin for decays
$B_d^0\rightarrow \eta\eta,\eta\eta',\eta'\eta' $ is very large:
$\varphi_{\eta\eta}^0(|P|/|T|=0.35, |S|/|P|=0.6)
-\varphi_{\eta\eta}^0(|P|/|T|=0.35, S=0)=-64^o(+360^o)$,
$\varphi_{\eta\eta'}^0(|P|/|T|=0.35, |S|/|P|=0.6)
-\varphi_{\eta\eta'}^0(|P|/|T|=0.35, S=0)=-76^o(+360^o)$,
$\varphi_{\eta'\eta'}(|P|/|T|=0.35, |S|/|P|=0.6)
-\varphi_{\eta'\eta'}(|P|/|T|=0.35, S=0)=35^o$. The singlet
penguin is very important for channels which contain the $\eta$
and  $\eta'$ mesons.

\section{$CP$ Violation}
It is interesting to calculate the $CP$-violation effects in our
model. $CP$-violating asymmetries for the decays of a neutral
$B_d$ into final states $\pi^+\pi^-$ and $\pi^0\pi^0$ are defined
as
\begin{equation}
\mathcal{A}_{\pi^+\pi^-}=\frac{|\langle
\pi^+\pi^-|W|\overline{B}_d^0\rangle|^2-| \langle
\pi^+\pi^-|W|B_d^0\rangle|^2}{|\langle
\pi^+\pi^-|W|\overline{B}_d^0\rangle|^2+| \langle
\pi^+\pi^-|W|B_d^0\rangle|^2}
\end{equation}
and
\begin{equation}
\mathcal{A}_{\pi^0\pi^0}=\frac{|\langle
\pi^0\pi^0|W|\overline{B}_d^0\rangle|^2-| \langle
\pi^0\pi^0|W|B_d^0\rangle|^2}{|\langle
\pi^0\pi^0|W|\overline{B}_d^0\rangle|^2+| \langle
\pi^0\pi^0|W|B_d^0\rangle|^2},
\end{equation}
with
\begin{eqnarray}
\langle \pi^+\pi^-|W|B_d^0\rangle&=&\sqrt{\frac{2}{3}}\langle (\pi\pi)_0|W|B_d^0\rangle+\frac{1}{\sqrt{3}}\langle (\pi\pi)_2|W|B_d^0\rangle,\\
\langle \pi^0\pi^0|W|B_d^0\rangle&=&-\frac{1}{\sqrt{3}}\langle
(\pi\pi)_0|W|B_d^0\rangle+\sqrt{\frac{2}{3}}\langle
(\pi\pi)_2|W|B_d^0\rangle.
\end{eqnarray}

\begin{figure}[h]
\begin{center}
\includegraphics[width=0.45\textwidth]{{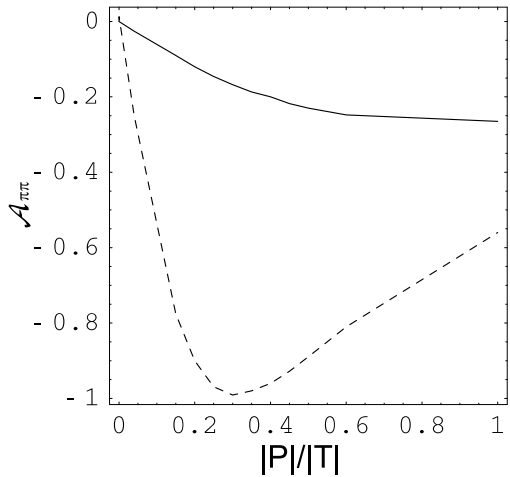}}
\end{center}
\caption{\small Influence of penguin contribution on $CP$
asymmetry in decays $B_d^0\rightarrow\pi^+\pi^-$ (solid line), and
$B_d^0\rightarrow\pi^0\pi^0$ (dashed line).
\label{asym}}
\end{figure}

$CP$ violation is still one of the least tested aspects of the
Standard Model. Current data exhibit $CP$ violation in the $B_d$
sector with large errors \cite{Au02}-\cite{CLEO}. In
Fig.~\ref{asym} we show the dependence of $CP$ asymmetry on the
size of penguin contribution. For small $\mathcal{A}_{\pi\pi}$ the
ratio $|P|/|T|$ should be very small. $CP$ violation effects are
more pronounced in the $\pi^0\pi^0$ channel, for example for
$|P|/|T|=0.35$ we have $\mathcal{A}_{\pi^+\pi^-}=-0.19$ , and
$\mathcal{A}_{\pi^0\pi^0}=-0.99$. Large values of these $CP$
asymmetries were obtained in other papers as well \cite{Xing}
\cite{Chau}. In our case, for $|P|$ comparable to $|T|$ the large
size of predicted $CP$ asymmetry is permitted by fairly large
FSI-induced phase shifts.

$CP$ asymmetries depend on strong phases
($\delta_T,\delta_C,\delta_P$, and $\delta_S$) of short-distance
amplitudes (\ref{4diagrams}). We know nothing about the size of
these parameters. In \cite{G&Rosner02} it is shown that for the
current data these phases may be in the region ($-90^o,90^o$)
\cite{Au02}-\cite{CLEO}. In order to show how these phases may
affect the calculations a few arbitrary phases where chosen. The
results  are given in Table~\ref{TAB4}. We assume that
$\delta_T=\delta_C=\delta_{TC}$ and
$\delta_P=\delta_S=\delta_{PS}$. From Table~\ref{TAB4} we see that
$CP$ asymmetries (for $|P|/|T|=0.35$) depend very strongly on
these phases, for example: $\mathcal{A}_{\pi^0\pi^0}\approx-1$
when we neglect short-distance amplitude phases, but
$\mathcal{A}_{\pi^0\pi^0}=-0.58 $  for $\delta_{TC}=30^o $ and
$\delta_{PS}=-20^o$. As shown in Table~\ref{TAB4}
 the $CP$ asymmetries  do not
depend significantly on $\epsilon$ (Eq.~\ref{epsilon}), so we may
keep $SU(3)$ symmetry in matrices $\textbf{U}$ and $\textbf{C}$
when analyzing $CP$ violation. The origin of big $CP$ asymmetry
lies in the joint effect of weak phases $\gamma$ and $\beta$ and
strong phases from inelastic rescattering, and short distance
amplitudes. Effects from FSI and short-distance amplitude mix.
Both effects give important contributions to $CP$ asymmetry.

\begin{table}[h]
\caption{\small Influence of strong phases $\delta_{TC}$ and
$\delta_{PS}$ of $T, C, P$, and $S$ amplitudes on $CP$ asymmetries
in decays $B_d^0\rightarrow\pi\pi$, for $|P|/|T|=0.35$. }
\label{TAB4}
\begin{center}
{ \footnotesize
\begin{tabular}{|c|c|| c c c|}
\hline $\mathcal{A}_{\pi\pi}$ & $\epsilon$&
\multicolumn{3}{c|}{Strong phases $\delta_{TC}$,
$\delta_{PS}=$}\\
& Eq.~\ref{epsilon}&$0^o,0^o$& $30^o,-20^o$ &$-20^o,30^o$ \\
 \hline \hline
$\mathcal{A}_{\pi^+\pi^-}$&$0.5 e^{-i36^o}$& -0.19  & 0.33  & -0.60 \\
& 1 &-0.19  & 0.30  & -0.57\\
\hline\hline
$\mathcal{A}_{\pi^0\pi^0}$&$0.5 e^{-i36^o}$     & -0.99    &-0.58   & -0.70 \\
& 1   & -1    &-0.63   & -0.65 \\
\hline
\end{tabular}
}
\end{center}
\end{table}
One has to realize, that for l=0 partial wave amplitudes the Regge
pole methods need not be reliable \cite{Rosner} and, consequently,
the obtained numbers should be considered as rough estimates only.
In additional to FSI effects and hadronic phases of 'bare' weak
diagrams, $CP$ violation effects may strongly depend on
electroweak diagrams~\cite{Rosner}, but there is not enough data
to determine the corresponding parameters.

\section{Conclusions}
In summary, we have discussed the effect of abandoning exact
$SU(3)$ in coupled-channel final-state interactions through
Reggeon exchange for $B^0_d\rightarrow \pi\pi , K\overline{K}$.
$SU(3)$ was broken by admitting lower lying trajectories for
strange Reggeons $|\epsilon|<1$ (Eq.~(\ref{epsilon})) and in the
Pomeron (Eq.~(\ref{pomeron})) couplings. As expected
in~\cite{Zen99} the singlet penguin diagram may be neglected in
intermediate states of $B^0_d\rightarrow\pi\pi,K\overline{K}$
decays. However, it cannot be neglected in decays to $\eta\eta$,
$\eta\eta'$, and $\eta'\eta'$. We have shown that strong FSI play
an important role in the analysis of $CP$-violating effects in $B$
decays. The size of $CP$ asymmetry in $B^0_d\rightarrow\pi\pi$
decays has been shown to depend strongly on the ratio of penguin
to tree amplitude and on the strong phases of short-distance quark
diagrams.

\medskip
\medskip
 {\large{\bf Acknowledgment}}

I would like to thank P. \.Zenczykowski for discussions and
comments regarding the presentation of the material of this paper.

\end{document}